\documentclass{PoS}

\title{Long distance contribution to $K_{L}$-$K_{S}$ mass difference}

\ShortTitle{Long distance contribution to $K_{L}$-$K_{S}$ mass difference}

\author{\speaker{Jianglei Yu}\\
       Department of Physics, Columbia University, New York, NY 10025, USA \\
       E-mail: \email{jy2379@columbia.edu}}

\abstract{We present a method to non-perturbatively determine the long-distance contribution to the $K_{L}$-$K_{S}$ mass difference. The calculation is performed on 2+1 flavor, domain wall fermion, $16^3\times32$ configurations with a 421 MeV pion mass and a kaon mass of 559 MeV . We include only current-current operators and drop all disconnected diagrams in the calculation. The largest contribution comes from quadratically divergent, short distance lattice artifacts. This quadratic divergence is eliminated through the GIM mechanism by introducing a valence charm quark. The remaining short distance effects are then removed by using RI/MOM technique which allows their exact replacement by the physical short distance part.}

\FullConference{XXIX International Symposium on Lattice Field Theory \\
		 July 10-16 2011\\
		 Squaw Valley, Lake Tahoe, California}

\begin{document}

\section{Introduction}
The extremely small $K_{L}$-$K_{S}$ mass difference has been measured accurately decades ago. The origin of this difference is the $K^0$-$\overline{K}^{0}$ mixing via second order weak interactions. Conventionally, the mass difference is separated into short-distance part and long-distance part. While the short-distance contribution has been calculated to next-to-leading order \cite{Herrlich:1993yv}, the long-distance contribution could only be determined non-perturbatively, which contributes around $30\%$ to the mass difference \cite{Buchalla:1995vs}. Norman Christ suggested a  Lattice QCD method to compute the long-distance contribution \cite{Christ:2010zz}. This proceeding is the first numerical experiment about the new method.


\section{Second order correlator}

\begin{figure}[htbp]
   \centering
   \includegraphics[width=0.7\textwidth]{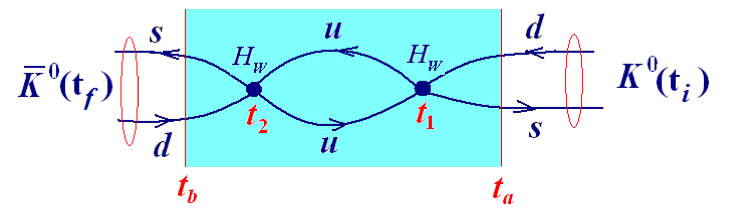} 
   \caption{Demonstration of the second order correlator  $\cal A$ in Equation \protect\ref{eq:amplitude}. Here the two four quark operators are integrated over the shadowed region.}
    \label{fig:schematic}
\end{figure}

To compute $K_L$-$K_S$ mass difference on a Euclidean space lattice, we could evaluate the time-integrated second-order product over a time interval $[t_a, t_b]$:
\begin{equation}
{\cal A}=\frac{1}{2}\sum_{t_1=t_a}^{t_b}\sum_{t_2=t_a}^{t_b}<\overline{K}^{0}(t_f)H_W(t_2)H_W(t_1)K^{0\dag}(t_1)>
\label{eq:amplitude}
\end{equation}
Here the Kaon is created at $t_i$, the second-order weak Hamiltonian acts between time interval $[t_a, t_b]$, and the outcome anti-Kaon is annihilated at $t_f$. The amplitude is represented schematically in Figure \ref{fig:schematic}. 

If we assuming that $t_a-t_i$ and $t_f-t_b$ are large enough for the interpolating operators to project onto Kaon states, after inserting a complete set of intermediate states, Eq \ref{eq:amplitude} becomes:
 \begin{equation}
{\cal A} = |Z_K|^2e^{-M_K(t_f-t_i)}\sum_{n}\frac{<\overline{K}^0|H_W|n><n|H_W|K^0>}{(M_K-E_n)^2}\left\{e^{(M_K-M_n)T}-(M_K-M_n)T-1\right\}
\label{eq:integrated_amp}
\end{equation}
Here $T=t_b-t_a+1$ and $Z_K$ is the normalization factor of Kaon interpolating operator. We assume that there is no intermediate state degenerate with kaon in this expression, which is true in this work. The term proportional to $T$ in Equation \ref{eq:integrated_amp} gives the finite volume approximation to $K_L$-$K_S$ mass difference.
\begin{equation}
\Delta M_K^{FV} =  2\sum_{n}\frac{<\overline{K}^0|H_W|n><n|H_W|K^0>}{M_K-E_n}
\end{equation}
The other terms in Equation \ref{eq:integrated_amp} on $T$ could be classified into three categories: i). The exponential decreasing term, if $E_n>M_K$, these terms could be neglected when $T$ is sufficiently large; ii). Exponential increasing term, if $E_n<M_K$, these terms must be identified independently and subtracted from the result; iii) The term independent of $T$, which is trivial.

\begin{figure}[htbp]
	\centering
	\includegraphics[width=0.7\textwidth]{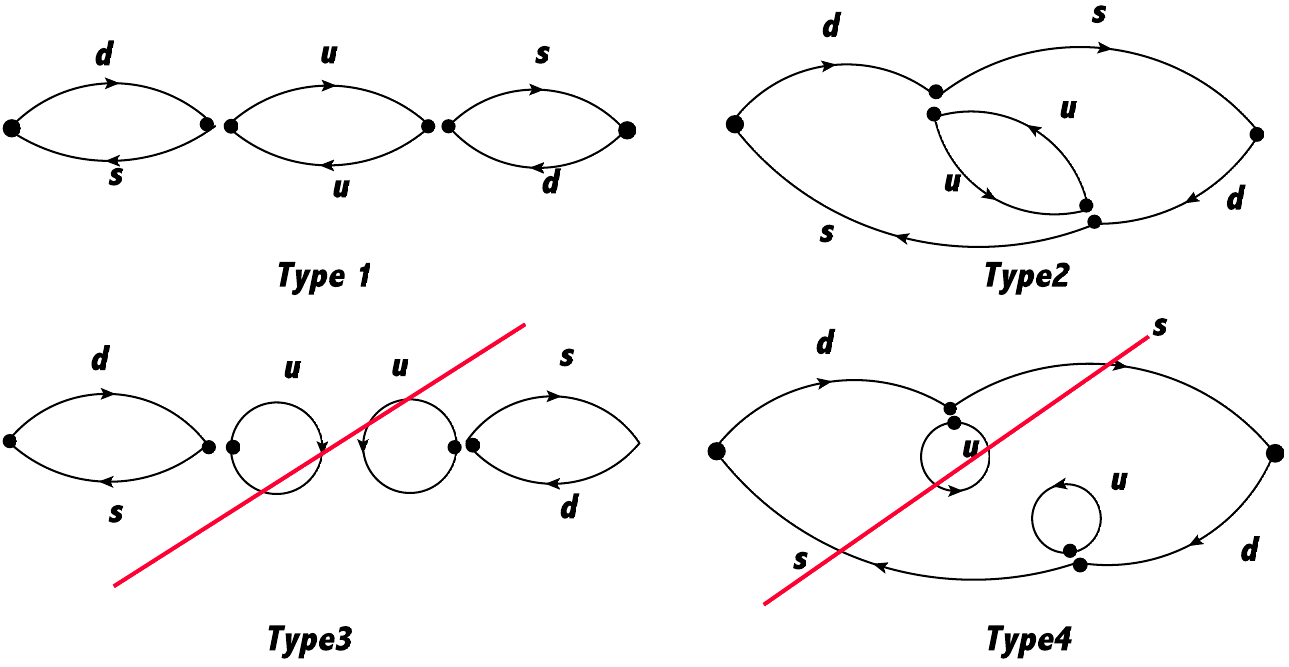}
	\caption{Four types of contractions in the  4-point correlator, only type 1 and type 2 are included in the calculation, type 3 and type 4 are dropped.}
	\label{fig:contraction}
\end{figure}

The full $\Delta S=1$ effective Hamiltonian consist of 7 independent four-quark operators \cite{Blum:2001xb}, we include only the current-current operator $Q_1$ in this work. 
\begin{equation}
Q_1 = (\bar{s}_\alpha d_\alpha)_{V-A}(\bar{u}_\beta u_\beta)_{V-A}
\label{eq:operator}
\end{equation}
The four different types of contractions are listed in Figure \ref{fig:contraction}. In this work, we only compute type 1 and type 2 contractions and drop type 3 and type 4 contraction. Type 3 contraction is dropped because it is disconnected diagram. We need to compute extra random source propagators for type 4 contraction. So we also drop type 4 contraction in this first numerical experiment. 

\begin{figure}[htbp]
	\centering
	\includegraphics[width=0.7\textwidth]{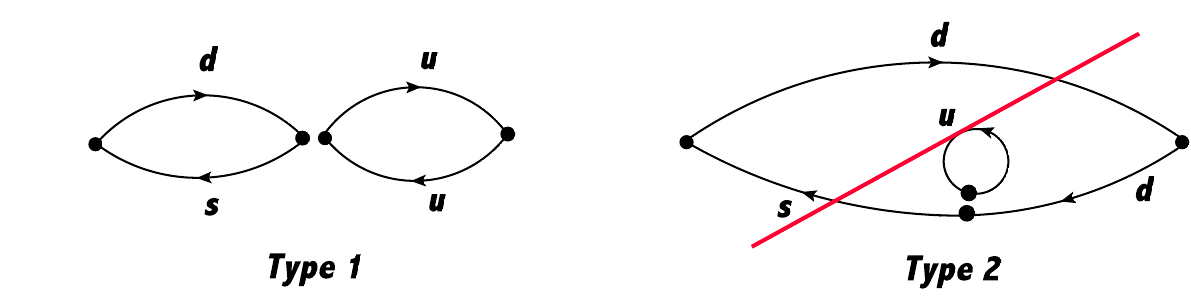}
	\caption{Two types of contractions in $<\pi^0|Q_1|K^0>$ 3-point correlator, type 2 is dropped to be consistent with 4-point correlator in Figure \protect\ref{fig:contraction}.}
	\label{fig:ktopi}
\end{figure}

As we mentioned before, we must identified the exponential increasing terms in Eq \ref{eq:integrated_amp}. These terms comes from the intermediate states which are lighter than Kaon, such states are $\pi^0$ state and vacuum state in this calculation. Since the disconnected diagrams are neglected, there will be no vacuum intermediate state. Then we must calculate $<\pi^0|Q_1|K^0>$ and subtract the exponential increasing contribution from Equation \ref{eq:integrated_amp}. The contractions in this calculation are given in Figure \ref{fig:ktopi}. We drop type 2 to be consistent with 4-point correlator calculation.

\section{Short distance effect}
\begin{figure}[htbp]
	\centering
	\includegraphics[width=0.6\textwidth]{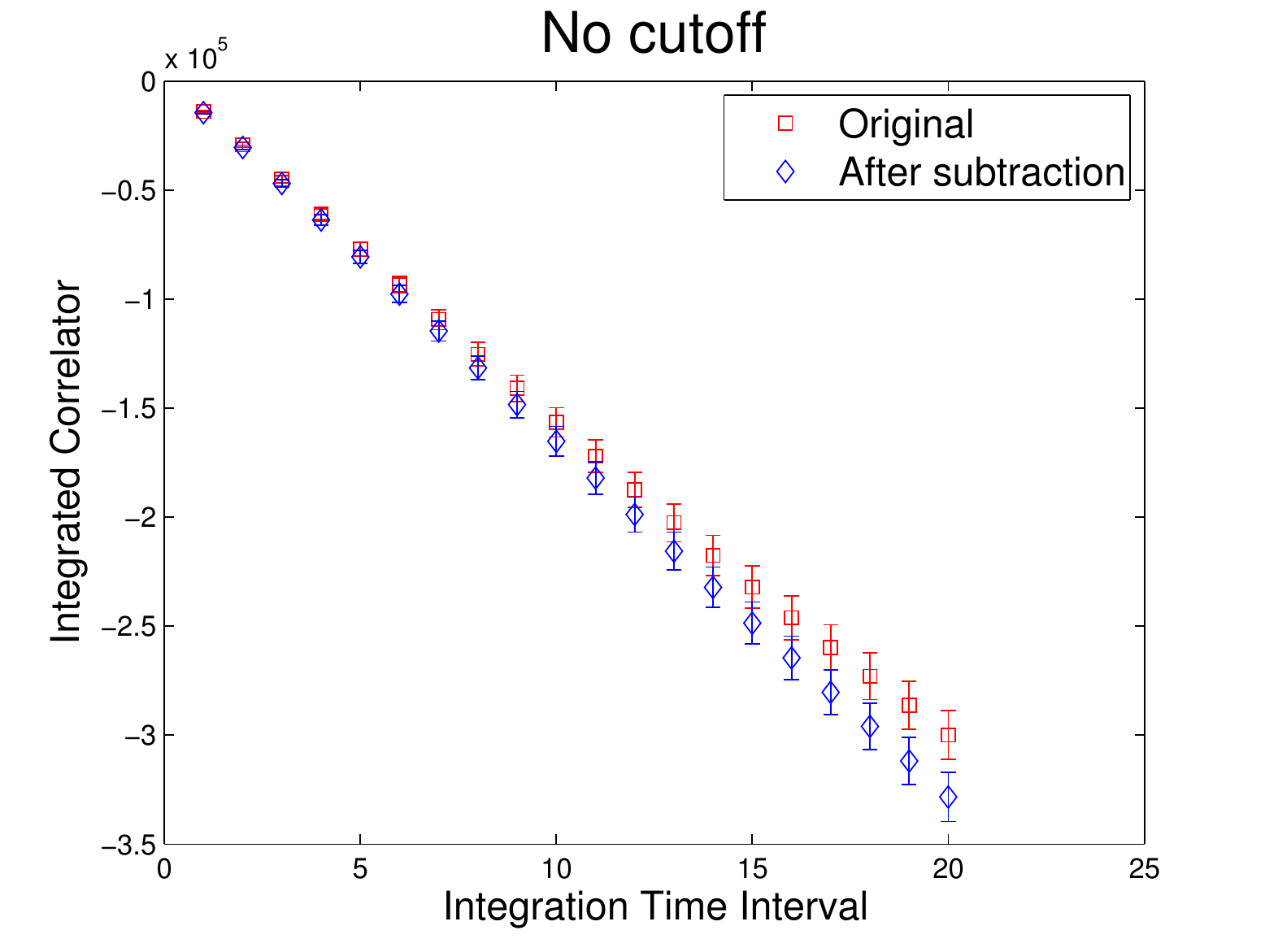}
	\includegraphics[width=0.6\textwidth]{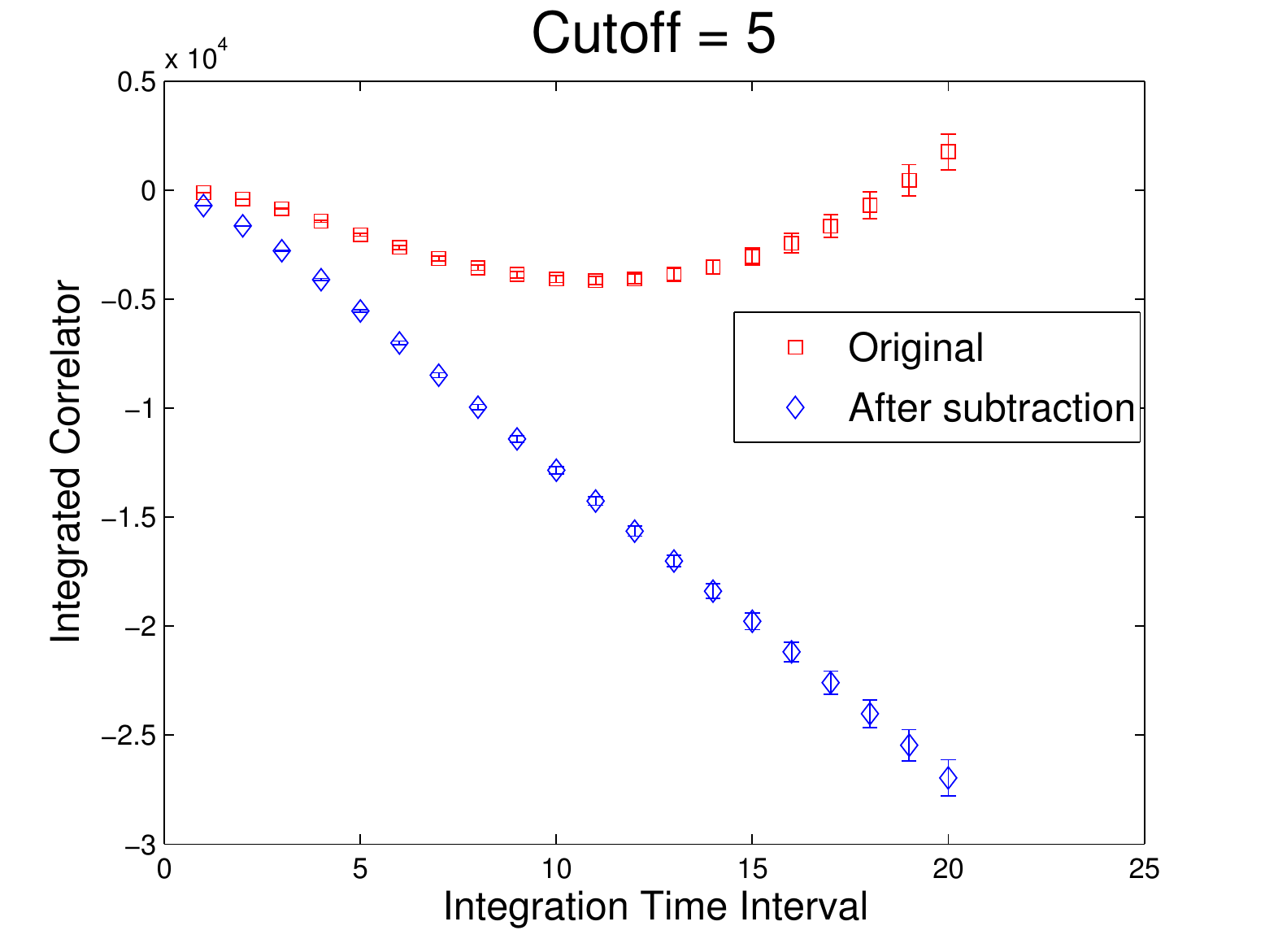}
	\caption{Integrated second order correlator as a function of integration time interval. Red and blue curve show the results before and after the subtraction of $\pi^0$ exponential term. First plot is the result without any cutoff. Second plot shows the result with cutoffs radius 5.}
	\label{fig:cutoff}
\end{figure}

The computation was performed on $N_f=2+1$ flavor $16^2\times32\times16$ lattice with DWF, Iwasaki gauge action, $a^{-1}=1.73(3)$ Gev, $421$ Mev pion mass and $559$  Mev kaon mass. The two wall-source kaons are located at time slice $t_i=0$ and $t_f=27$. The two $\Delta S=1$ operator act between time slice $[4,23]$. We calculated 600 configurations separated by 10 Monte Carlo time units. The result is given in the first plot in Figure \ref{fig:cutoff}. The plot shows the integrated second order correlator as a function of integration time interval T. For each given integration time interval T, we calculate the integrate correlator in Equation \ref{eq:integrated_amp} between all possible time interval $[t_a, t_a+T-1]$ and take the average as final result. The two curves in the plot are the results before and after the subtraction of $\pi^0$ exponential term. The results have both long-distance part and short-distance part. The short distance part means that the two $\Delta S=1$ operator are close to each other on the lattice. We expect the short distance part to be quadratically divergent because of the up quark loop in Figure \ref{fig:schematic}. To get a detailed understanding of the short distance effect, we could introduce an artificial cut off, i.e., require the separation between two operators $|x_2-x_1|> r$. This cutoff will reduce the short distance effect and the long distance part will remain untouched. The other plot in Figure \ref{fig:cutoff} show the results with cutoff radius 5. We can see that the amplitude of the result is reduced substantially after introcuding the cutoff. And the contribution from $\pi^0$ intermediated state becomes visible. We could measure the mass difference at different cutoffs. From Equation \ref{eq:integrated_amp}, the mass difference is given by the coefficient of the linear term up to some factor when T is large. We choose to fit the slope of integrated correlator plot in the range $T\in[11,20]$. The mass differences are listed in Table \ref{tab:cutoff}. We could do a naive inverse quadratic fit for the mass differences at different cutoffs. The result is in Table \ref{tab:cutoff}, which suggest the short distance effect is quadratically divergent. 
\begin{table}[ht]
\caption{Mass differences at different cutoff radius}
\label{tab:cutoff}
\begin{center}
\begin{tabular}{cccccc}
\hline
Cutoff Radius & 1 & 2 & 3 & 4 & 5\\
\hline
$\Delta M_K$ & 0.3342(80) & 0.1533(30) & 0.0796(17) & 0.0560(15) & 0.0455(14) \\
\hline
\end{tabular}
\end{center}
\end{table}

\begin{figure}[htbp]
	\centering
	\includegraphics[width=0.6\textwidth]{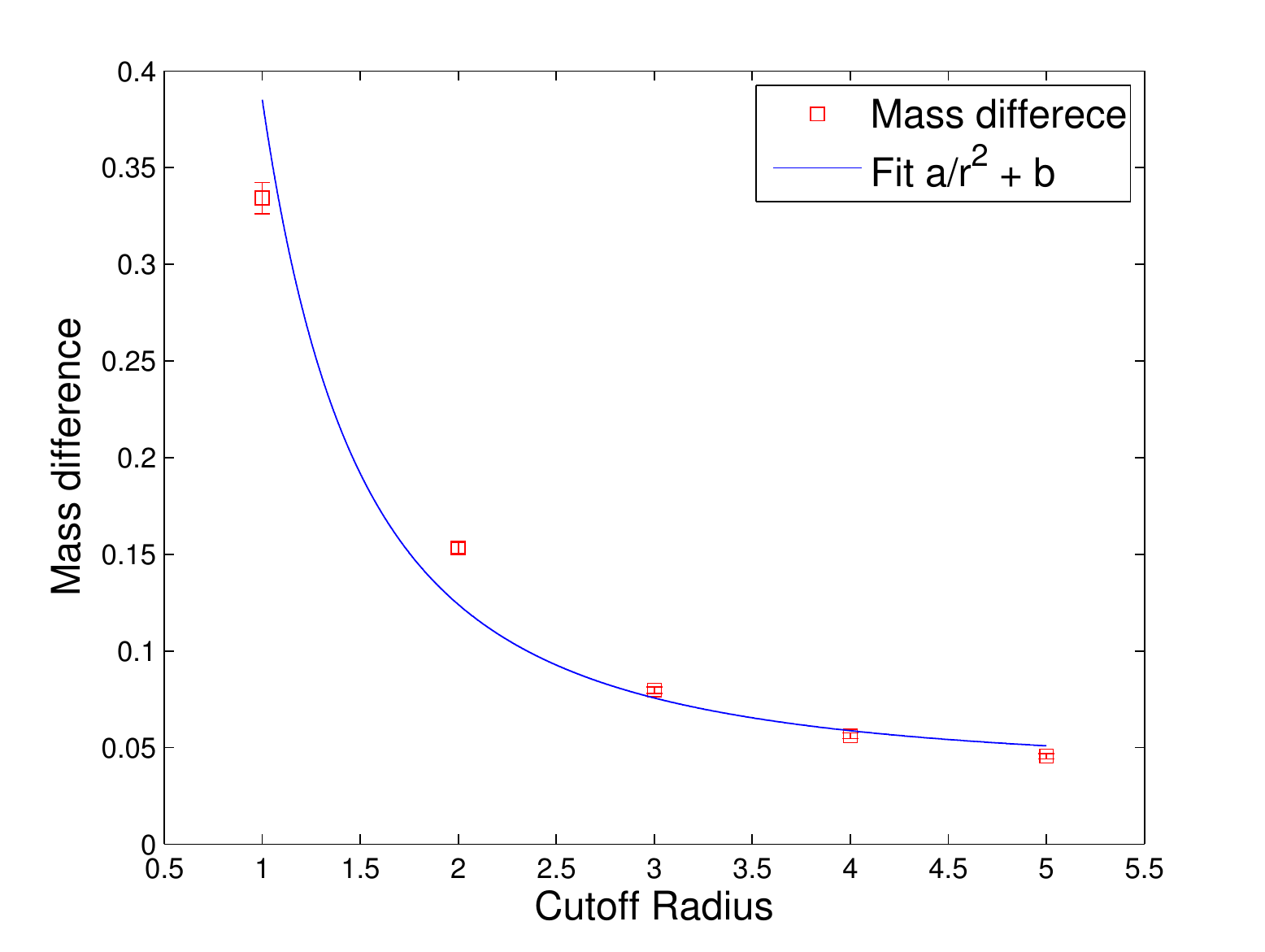}
	\caption{The mass differences at different cutoff radius, the blue cure is a naive two parameter fit.}
	\label{fig:mass_cutoff}
\end{figure}

\vspace{-5mm}
\section{Charm quark and GIM}
\begin{figure}[htbp]
	\centering
	\includegraphics[width=0.6\textwidth]{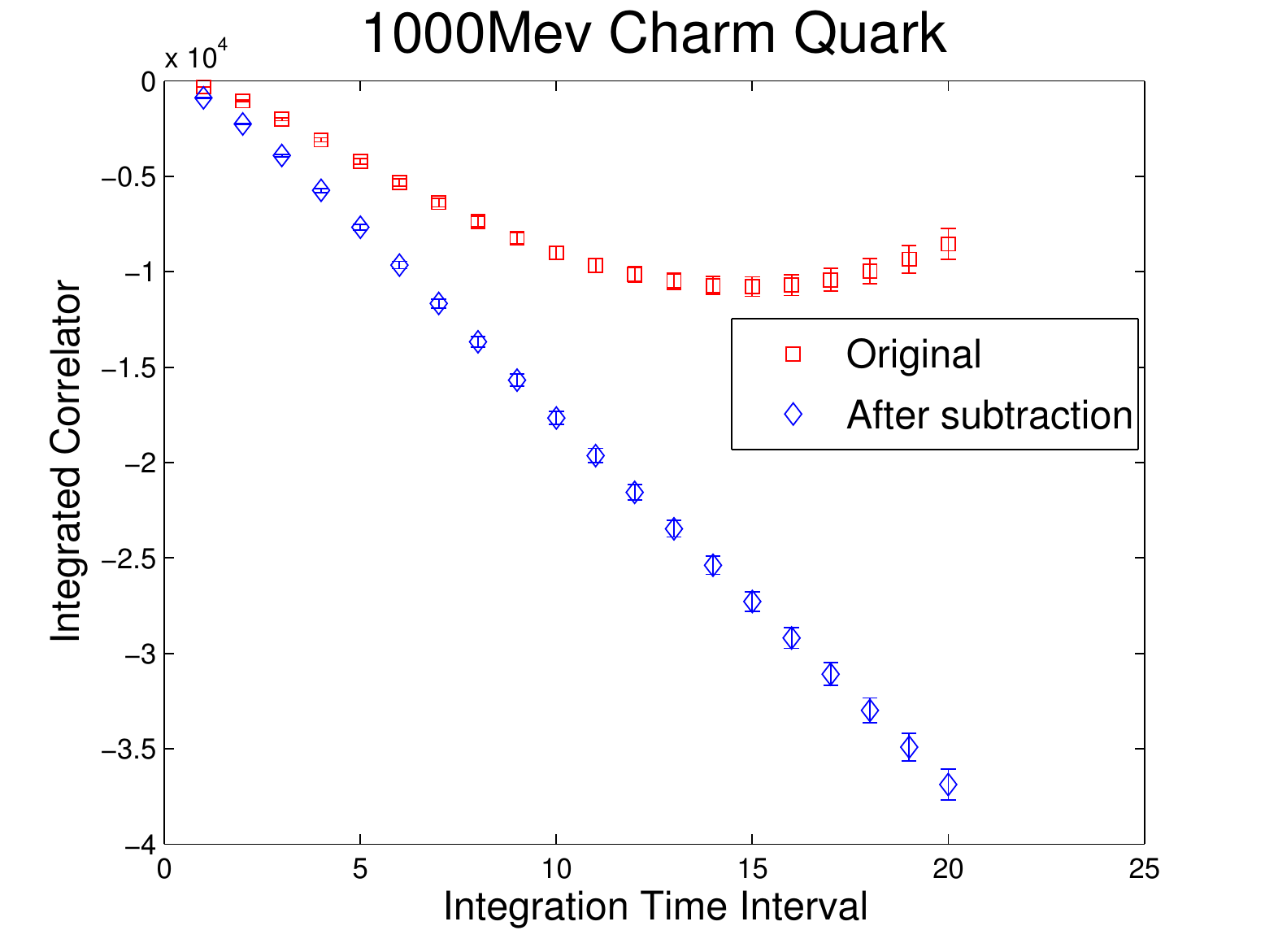}
	\caption{The integrated correlator after the inclusion of 1000Mev charm quark, Red and blue curve show the results before and after the subtraction of $\pi^0$ exponential term}
	\label{fig:charm}
\end{figure}

In order to remove the short distance, we introduce valence charm quark to the calculation. Then GIM mechanism will reduce the quadratic divergency into logarithmically. To implement this in the lattice calculation, we could replace all the up quark propagators in Figure \ref{fig:contraction} with the difference between up quark propagator and charm quark propagator. We use $5$ difference valence charm quark mass ranged from 200Mev  to 1000Mev. The integrated correlators after GIM subtraction are plotted in Figure \ref{fig:charm}. The mass difference for difference valence charm quarks are listed in Table \ref{tab:charm}.

\section{Remain Short distance effect}
\begin{figure}[htbp]
\centering
\includegraphics[width=0.7\textwidth]{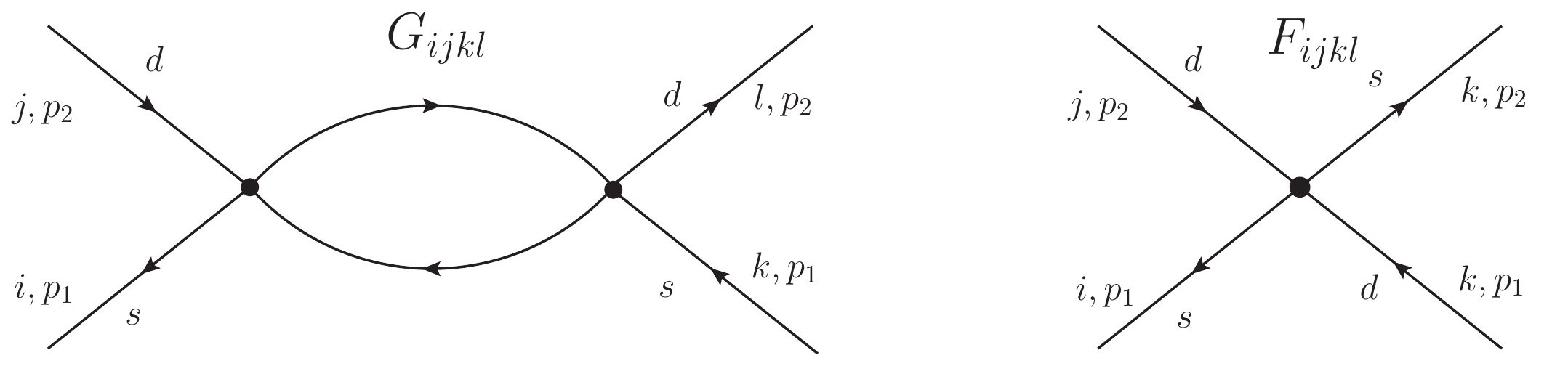}
\caption{Off-shell, amputated four quark green functions, left diagram is for two $\Delta S=1$ operators, right diagram is for one $\Delta S=2$ operator.}
\label{fig:npr_demo}
\end{figure}

Even with the valence charm quark, there will still be some short distance lattice artifacts remain. The short distance part in Equation \ref{eq:amplitude} could be given by:
\begin{eqnarray}
{\cal A}_{SD}&=&\frac{1}{2}\sum_{t=t_a}^{t_b}<\overline{K}^0(t_f)C(\mu){\cal O}(t)K^{0\dag}(t_i)>\nonumber \\
                       &=&\frac{1}{2}|Z_K|^2e^{-M_K(t_f-t_i)}C(\mu)<\overline{K}^0|{\cal O}|K^0>T
\end{eqnarray}
Here $T=t_b-t_a+1$, ${\cal O}=(\bar{s}d)_{V-A}(\bar{s}d)_{V-A}$, $C(\mu)$ is the conversion factor at a certain momentum scale $\mu$. To identify $C(\mu)$ by using RI/SMOM technic, we could evaluate off-shell, four quark, amputated green function for the two diagrams in Figure \ref{fig:npr_demo} at some large external momentum scale $\mu$. 
The external momentum satisfy $p_1^2=p_2^2=(p_1-p_2)^2=\mu^2$. Suppose the result for two diagrams are $G_{ijkl}$ and $F_{ijkl}$ respectively. Then we project the green functions into the desired gamma structure $P_{ijkl}=((1-\gamma^5)\gamma_{\mu})_{ji}((1-\gamma^5)\gamma_{\mu})_{lk}$. The conversion factor is given by Equation \ref{eq:npr}. 
\begin{equation}
G(\mu)=G_{ijkl}P_{ijkl} \quad F(\mu)=F_{ijkl}P_{ijkl} \quad C(\mu) = \frac{G(\mu)}{F(\mu)}
\label{eq:npr}
\end{equation}

\begin{figure}
\centering
\includegraphics[width=0.5\textwidth]{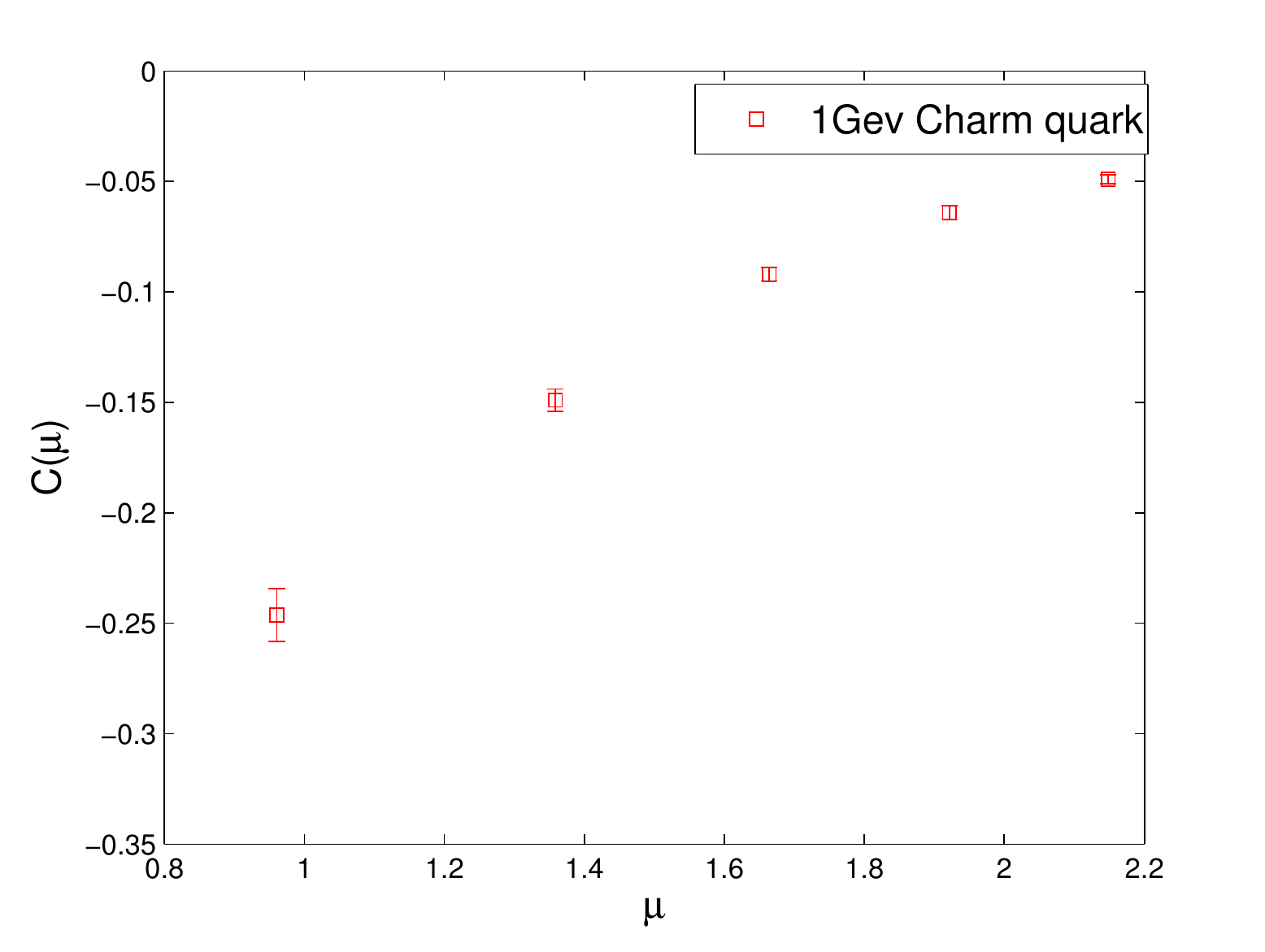}
\hspace{-5mm}
\includegraphics[width=0.5\textwidth]{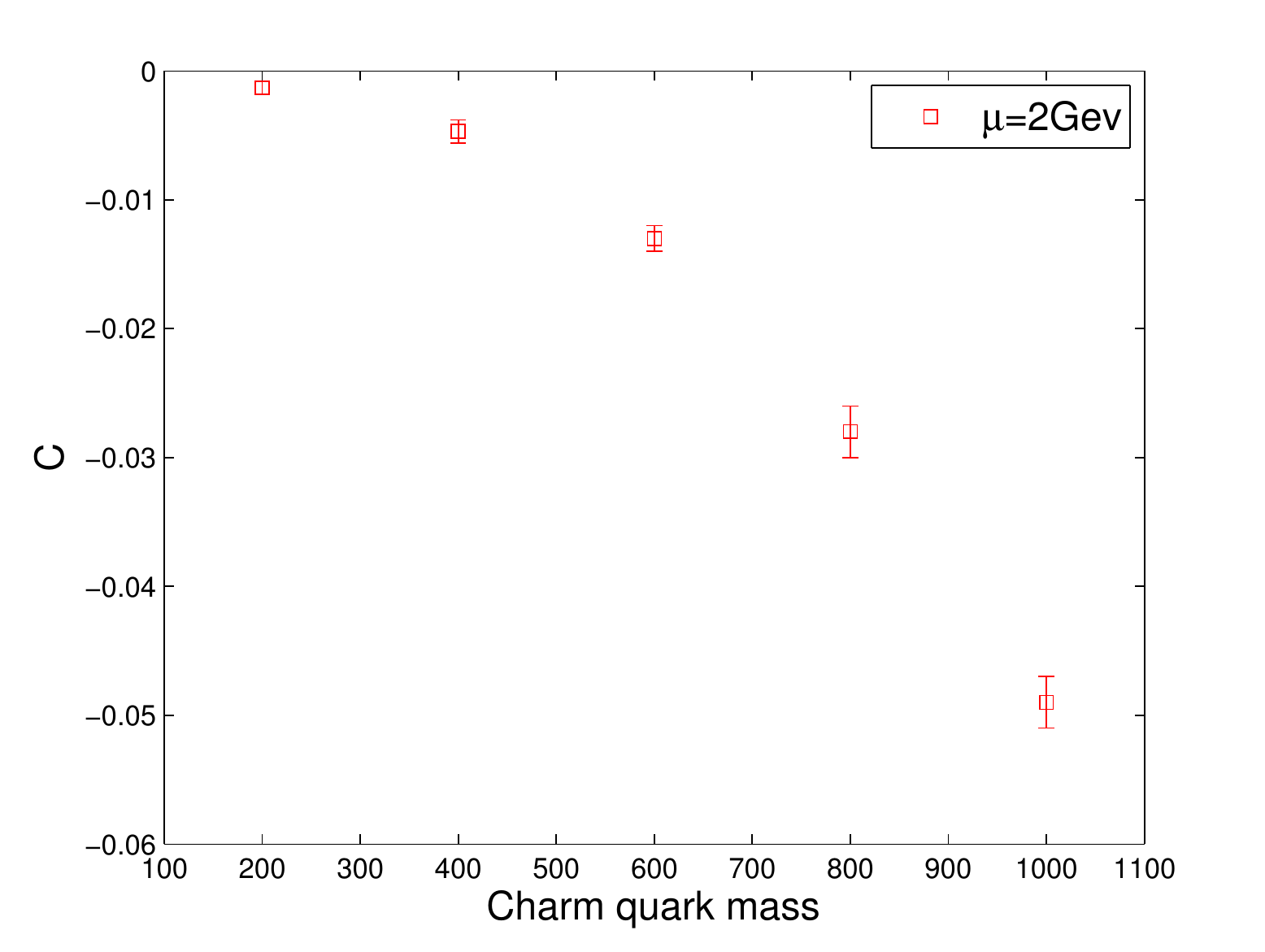}
\label{fig:npr_result}
\caption{The left plot shows the conversion factor $C(\mu)$ as a function of momentum scale qith 1 Gev charm quark. The right plot shows $C(\mu)$ at different charm quark masses at fixed $\mu$ = 2Gev.}
\end{figure}

In Figure \ref{fig:npr_result}, the left plot shows the dependence of $C(\mu)$ on the momentum scale $\mu$ if we use 1 Gev valence charm quark. As we expected, $C(\mu)$ will decrease while increasing $\mu$, because the difference between charm quark and up quark will decrease while the momentum scale get larger. In the right plot, we fix momentum scale to be 2 Gev and plot $C$ as a function of charm quark mass. When charm quark mass get smaller, the remain short distance effect becomes smaller. In Table \ref{tab:charm}, we show the remain short distance effect  at different charm quark masses. We conclude that the remain short distance effects are so small that we could neglect them in this work.

\begin{table}[ht]
\caption{Mass differences at different valence charm quark masses}
\label{tab:charm}
\begin{center}
\begin{tabular}{cccccc}
\hline
$M_c$(Mev) & 200 & 400 & 600 & 800 & 1000\\
\hline
$\Delta M_K$ & 0.0440(10) & 0.0455(12) & 0.0496(13) & 0.0556(14) & 0.0628(15) \\
\hline
$(\Delta M_K)_{SD}$ & 6.2e-5 &  2.4e-5 & 6.2e-4 & 0.0013 & 0.0023\\
\hline 
\end{tabular}
\end{center}
\end{table}

\section{Conclusion}
We perform a numerical study of the long distance part of $K_L$-$K_S$ mass difference. The short distance part could be reduce by the inclusion of valence charm quark. The exponential increasing term could be identified and subtracted. The remain short distance effect could be computed by using RI/SMOM technic and removed from the results.

The author thanks all his RBC/UKQCD collaborators for discussions and suggestions. Especially thanks to Prof. Norman Chris for detailed instructions and discussion. 

\bibliography{citations}
\bibliographystyle{JHEP}
\end{document}